\documentclass[11pt]{article}
\usepackage{times}
\usepackage{cospar}
\usepackage{url}
\usepackage{epsfig} 
\usepackage[sectionbib]{natbib}
\renewcommand{\bibsection}{\section*{REFERENCES}}
\renewcommand{\bibhang}{\setlength{8mm}}
\renewcommand{\bibsep}{\setlength{0mm}}
\usepackage{graphicx}

\def\apj{{\em Astrophys. J}}                          
\def\apjl{{\em Astrophys. J}}                         
\def\mnras{{\em Mon. Not. R. Astron. Soc.}}           
\def\aap{{\em Astron. Astr.}}                         
\def\asr{{\em Adv. Space Res.}}                       
\def\apss{{\em Adv. Space Sci.}}                      
\def\prd{{\em Phys. Rev. D}}                          
\def\ssr{{\em Space Sci. Rev.}}                       
\def\teq#1{$\, #1\,$}                           
{\catcode`\@=11                                                   
\gdef\SchlangeUnter#1#2{\lower2pt\vbox{\baselineskip 0pt\lineskip0pt     
\ialign{$\m@th#1\hfil##\hfil$\crcr#2\crcr\sim\crcr}}}}            
\def\gtrsim{\mathrel{\mathpalette\SchlangeUnter>}}                
\def\lesssim{\mathrel{\mathpalette\SchlangeUnter<}}  

\font\sixrm=cmr6 scaled \magstep0 
\def\dover#1#2{\hbox{${{\displaystyle#1 \vphantom{(} }\over{ 
   \displaystyle #2 \vphantom{(} }}$}}         
\def\erg{\varepsilon}     
 
\def\emax{\erg_{\hbox{\sixrm MAX}}} 
\def\emin{\erg_{\hbox{\sixrm MIN}}} 
\def\gammamin{\gamma_{\hbox{\sixrm MIN}}} 
\def\dpsr{\hbox{$d_{\hbox{\sixrm PSR}}$}} 
\def\dpsrsq{\hbox{$d^2_{\hbox{\sixrm PSR}}$}} 
\def\ethresh{\erg_{\hbox{\sixrm TH}}} 
\title{HIGH-ENERGY EMISSION FROM PULSARS:\\ THE POLAR CAP SCENARIO}

\author{Matthew G. Baring\address{Rice University,
Dept. of Physics and Astronomy, MS-108,\\
P. O. Box 1892, Houston, TX 77251-1892}} 

\begin{document}

\newcommand{\vol}[3]{$\,$ \bf #1\rm , #2, $\,$ #3.}         

\maketitle

\vphantom{p}
\vskip -190pt
\centerline{\hfill Adv. Space Research, in press}
\vskip 165pt

\begin{abstract}
The study of pulsars in the three and a half decades since their
discovery has highlighted a handful of issues critical to their
understanding.  To date there is no consensus on the physical mechanism
for their radio radio emission, despite a rapid increase in the
observed population due to the Parkes Multi-Beam survey and prospects
for similar growth in the radio population database in the near
future.  The small subset of pulsars that emit at X-ray to gamma-ray
wavelengths are critical to refining the pulsar paradigm since this
energy band (i) is where the vast majority of radiative luminosity is
observed, and (ii) is intimately connected to the pair winds that form
the dominant mode of energy deposition in the circum-pulsar
environment.  The most crucial point of contention pertaining to the
high energy astrophysics of pulsars is the location of the acceleration
region in their magnetospheres:  is an outer gap model or a polar cap
scenario (or both) the most appropriate picture.  Radiative signatures
provide the clues to this current enigma.  This review focuses on
salient characteristics of the polar cap scenario; these form the basis
for discriminating observational diagnostics that should drive pulsars
studies in the GLAST era just three years away.
\end{abstract}

\section*{INTRODUCTION}

Pulsar astronomy is at a fascinating juncture with many exciting new
results from the Chandra and XMM X-ray telescopes, and the prospects of
next generation gamma-ray experiments, led by space missions INTEGRAL
(just launched), GLAST and AGILE, and a number of atmospheric Cerenkov
telescopes coming on line in the next few years.   These will enable a
continuation of ground-breaking discoveries.  Pulsars are expected to
be detected by GLAST (Gamma-Ray Large Area Space Telescope: {\tt
http://www-glast.stanford.edu}) in profusion, some that are
radio-selected, like most of the present EGRET/Comptel pulsars, and
perhaps even more that are detected via independent pulsation
searches.  This theoretical review summarizes relevant characteristics
of the polar cap model, emphasizing distinctions from the competing
outer gap model.  These features include acceleration properties, the
X-ray to gamma-ray spectral shape, high energy cutoffs, pulse profiles
and flux observabilities in different wavebands, and how these
characteristics generally depend on pulsar period and period
derivative.  The polar cap scenario exhibits definitive signatures that
will be readily tested by the detections of GLAST and other
experiments, thereby establishing cogent observational diagnostics.

This polar cap review complements the outer gap review by Cheng (2003).
The key distinguishing characteristic of the two models that
is usually cited is that the acceleration zone for the polar cap model
is confined to within a few stellar radii of the pulsar surface, while
the region of acceleration in the outer gap model is proximate to the
light cylinder.  However, in terms of the physical manifestations that
are most directly probed by astronomical observations, the presence or
absence of an extremely strong magnetic field is the conspicuous
feature.  In polar cap models, the strong field permits single photon
pair creation that attenuates super-GeV photons in Crab and Vela-like
pulsars, whereas pair creation in outer gap models is mediated through
the more familiar two-photon process involving surface thermal X-rays
as targets.  It is a principal contention of this paper that the two
models will not generate coincident predictions for a large population
of pulsars (contrasting isolated sources) that sample a significant
range of periods and period derivatives.  Since such model
discrimination will probably be rendered by the gamma-ray observations
of GLAST, here the focus is on magnetospheric cascade spectral
properties.  The reader is referred to the review by Michel (2003), 
and Mestel (2000) for discussions of pulsar electrodynamics.
Recent polar cap model reviews include those of Harding (2001) and
Rudak, et al. (2002).  The issues of radio pulsar death lines
and radio quiescence at high fields are beyond the scope of this paper;
discussions can be found in Baring (2001a) and Zhang (2003).
 
\section*{POLAR CAP MODEL BASICS}

The critical physics ingredient for the polar cap model is the presence
of a strong magnetic field in the acceleration and cascade emission
region.  Such high fields follow from the contention that the induced
electric fields that seed particle acceleration in the oblique rotators
exists at low altitudes near the neutron star surface at the magnetic
poles.  Since the earliest polar cap pulsar models of Sturrock (1971)
and Ruderman and Sutherland (1975; hereafter RS75), there has been a
potpourri of variations and updates, with the primary division being
whether or not there is free emission of particles from the neutron
star surface.  Whether the surface temperature \teq{T} of the neutron
star exceeds the ion, \teq{T_{\rm i}} and electron, \teq{T_{\rm e}},
thermal emission temperatures controls the nature of the acceleration
zone.

If \teq{T < T_{\rm i}}, ions will be trapped in the neutron star crust
(RS75, Usov and Melrose 1995) and a vacuum gap will develop at the
surface forming the locale of the region of particle acceleration and
radiation.  If \teq{T > T_{\rm e}}, free emission of particles of
either sign of charge will occur.  The flow of particles is then
limited only by space charge, and since such particle flow all along
each open field line is unable to supply the corotation charge that is
required to short out the electric field component \teq{E_{\parallel}}
along the magnetic field lines, an accelerating potential will develop
(Arons and Scharlemann, 1979; Muslimov and Tsygan, 1992).  In space
charge-limited flow (SCLF) models, the accelerating \teq{E_{\parallel}}
is screened at a height where the particles radiate \teq{\gamma}-rays
that produce pairs.  This so-called pair formation front (e.g.  Arons,
1983; Harding and Muslimov, 1998; Muslimov and Harding, 2003) can occur
at high altitudes above the polar cap, depending on the colatitude of
the field line (discussed below in the section on Slot Gaps), a
property that may prove necessary to explain the spectral cutoffs in
the some or most of the EGRET pulsars.  Many surface temperatures have
now been measured for canonical X-ray pulsars in the range \teq{T \sim
10^5 - 10^6} K (Becker and Tr\"umper, 1997), though higher values are
obtained (\teq{T \sim 4\times 10^6 - 7\times 10^6} K: see Perna et al.,
2001) in observations of anomalous X-ray pulsars and soft gamma
repeaters, so both vacuum gaps and SCLF models need to be considered
depending on the source.

In the strong electric fields, the acceleration of primary electrons is
rapid and ceases when one of two types of radiative cooling becomes
significant, thereby establishing the maximum Lorentz factor
\teq{\gamma_e} of these particles, and generating a quasi-monoenergetic
primary distribution prior to cascading.  The cooling mechanisms are
curvature radiation (present in models from the earliest days of pulsar
theory; e.g. Sturrock, 1971) induced by the non-uniform magnetic field,
and resonant (magnetic) inverse Compton scattering of thermal X-rays
from the stellar surface (e.g. Sturner and Dermer, 1994), a more recent
incorporation.  Both are strong functions of the magnetic field
strength and either the electron's Lorentz factor or the field
geometry.  Curvature radiation-initiated cascades generally have
\teq{\gamma_e\sim 10^7} (e.g. Daugherty and Harding, 1989; see also
Harding and Muslimov, 1998), while inverse-Compton seeded pair cascades
yield \teq{\gamma_e\sim 3\times 10^5}--\teq{10^6} (e.g. Sturner, 1995;
see also Harding and Muslimov, 1998).

The primary photons propagate through the magnetosphere until they
achieve sufficient angles \teq{\theta_{\rm kB}} with respect to the
magnetic field to permit the creation of pairs via \teq{\gamma\to
e^+e^-} above the threshold energy of \teq{2m_ec^2/\sin\theta_{\rm
kB}}.  This propagation is impacted by general relativistic distortions
of photon trajectories and field structure (e.g. Gonthier and Harding,
1994; Harding, et al., 1997), as is the magnitude of the
field in the local inertial frame, so that curved spacetime properties
significantly modify the rates of pair creation.  For small polar cap
sizes, corresponding to longer pulsar periods, the primary photons fail
to acquire sufficient angles \teq{\theta_{\rm kB}} at low to moderate
altitudes prior to the decline of the dipole field, thereby permitting
the photons to escape unattenuated; pair creation is quenched at high
altitudes since the rate is a strongly increasing function of
\teq{B\sin\theta_{\rm kB}} (e.g. Tsai and Erber, 1974).  It is this
effect that is primarily responsible for the existence of a theoretical
death line for radio pulsars (Sturrock, et al. 1976) at longer
periods.  The inability of simpler curvature radiation-seeded cascades
to account for emission from pulsars of the longest periods has lead to
recent refinements (Zhang, et al., 2000; Hibschman and Arons,
2001; Harding, et al., 2002) that incorporate the influence
of non-resonant inverse Compton scattering of surface X-rays in seeding
cascades and subsequent pair creation for significantly smaller polar
cap sizes and small period derivatives.
 
The first generation of pair creation initiates the pair cascade, with
pairs generally being created in excited transverse (to the field)
momentum states, the so-called Landau levels.  Rapid de-excitation via
cyclotron and synchrotron radiation then follows (at least in pulsars
with Crab-like or lower fields), and these secondary photons travel to
higher altitudes and create further pairs and successive generations of
photons in a cascade.   The typical number of generations is around
3--4, and the total number of pairs per primary electron being of the
order of \teq{10^3}--\teq{10^4} (e.g. Daugherty and Harding, 1982).  The
cumulative product is an emission spectrum that comprises a
curvature/inverse Compton continuum that is cut off at hard gamma-ray
energies by pair creation, with the addition of several synchrotron
components at successively lower energies, terminating when the
magnetosphere becomes transparent to \teq{\gamma\to e^+e^-} at
significant altitudes.

A notable exception to this cascade scenario arises in
highly-magnetized pulsars such as PSR 1509-58, when the surface polar
field \teq{B_0} exceeds around \teq{6\times 10^{12}}Gauss.  Pairs are
then produced in the zeroth (ground state) Landau level (e.g. Baring and 
Harding, 2001; hereafter BH01), so that cyclotron/synchrotron emission
is prohibited.  Cascading is then effectively quenched at low altitudes
and the pair yield diminished.  While Zhang and Harding (2000)
suggested a possible alleviation of this cascade suppression via Landau
level excitation of higher generation pairs by Compton scatterings with
surface X-rays from the surface, Baring and Harding (2001) determined
that the population of excited Landau states relative to that in the
ground state is small, since such excitation can only achieved for photon
energies exceeding the cyclotron resonance (e.g. Gonthier et al., 2000).

Another profound alteration to the nature of cascades arises in high
field pulsars, due to the action of magnetic photon splitting,
\teq{\gamma\to\gamma\gamma}, a third-order quantum electrodynamical
process in which a single photon splits into two lower-energy photons
(e.g. Adler, 1971; Baring and Harding, 1997).  The rate of splitting, like
that of magnetic pair creation \teq{\gamma\to e^+e^-}, is generally a
rapidly increasing function of field strength (for fields
\teq{B\lesssim 10^{14}}Gauss), photon energy and photon propagation
angle with respect to the field.  Since splitting possesses no energy
threshold, it can dominate the first order process of pair creation if
{\bf B} is sufficiently high, typically above \teq{\sim 10^{13}}Gauss
(Harding, et al., 1997; hereafter HBG97).  This leads to an
alternative channel for cascade quenching, with gamma-rays being
reprocessed without yielding pairs so that synchrotron generations are
suppressed.  The operation of splitting produces distinctive spectral
bumps and polarization signals in the EGRET/Comptel band (HBG97);
splitting-influenced pulsar cascades are also addressed in depth in
Baring and Harding (2001).

\section*{SPECTRAL AND TEMPORAL SIGNATURES}  

Spectral and temporal properties can define the most potent means for
distinguishing between polar cap and outer gap scenarios.  Yet such
diagnostics are limited in power because of the significant number of
parameters in each of the models.  This degree of flexibility renders
each model capable of approximately describing the major
characteristics presented in a given dataset on an individual pulsar: a
classic example is the comparison of polar cap models (e.g. Daugherty
and Harding, 1996; Dyks and Rudak, 2000) with outer gap predictions 
(Romani, 1996) for the Vela pulsar.  Such a predictive redundancy of two
disparate models for individual sources is a limitation that is lifted
when global characteristics of the pulsar population are considered.
This provides motive for emphasizing global trends and properties for
the entire pulsar population in this paper.  The reasonable premise
underpinning this tack is that the two scenarios cannot generate
identical trends for luminosities, spectral indices, maximum gamma-ray
energies, non-thermal X-ray indices, pulse profiles, polarization
signatures, etc. as functions of period, period derivative, and viewing
perspective for dozens or hundreds of well-measured pulsars.  Since
such large gamma-ray pulsar databases will be afforded by the GLAST
mission, such an approach is both pertinent and timely.

\subsection*{Trends and Distinguishing Characteristics}

In the polar cap model, there are two distinguishing spectral features
that serve as indicators of the field strength, namely structure sampling
the cyclotron resonance that generally arises in the soft or hard X-ray
bands for Crab-like and Vela-like pulsars, and the super-GeV cutoff due
to attenuation by \teq{\gamma\to e^{\pm}} pair creation.  In principal,
these features are smeared out by distribution of the emission of
magnetospheric altitudes and colatitudes; in practice this spectral
structure degradation impacts the cyclotron feature much more than it
does the high energy cutoff.  Moreover, the cyclotronic structure is
blueshifted by the minimum Lorentz factor of pairs generated in a
synchrotron cascade: in most cases this further broadens the structure
rendering field diagnostics difficult if not impossible.  In addition,
structure in the primary curvature spectrum due to cooling breaks can
confuse the situation in the hard X-ray band.

Monte Carlo cascade simulations (e.g. Daugherty and Harding, 1982)
generate minimum Lorentz factors \teq{\gammamin} of around 50--100 for
secondary pairs in models of Vela-like pulsars.  The synchrotron photon
energy for this Lorentz factor is \teq{\emin\sim\gammamin B/B_{\rm
cr}}, noting that a factor of \teq{1/\gammamin} is introduced to
account for the cascade beaming angle.  This energy, the energy of cascade
cessation, is typically in hard X-rays for Crab-like pulsars, and
generally \teq{\gammamin} depends on \teq{B_0}, \teq{P}, etc. in more
or less the same manner (Baring and Harding, 2000) that \teq{\emax} does
in Eq.~(\ref{eq:emax}) below, since the same pair creation physics
applies to both.  The result is an \teq{\emin} with a fairly weak field
dependence.  Note, however, that the reprocessing that leads to the
establishment of \teq{\emin} is initiated at slightly lower altitudes
(corresponding to more cascade generations) than the ultimate
attenuation that defines \teq{\emax}, thereby complicating parametric
specifications of \teq{\emin}.  The spectrum possesses a break at
\teq{\emin}, below which it assumes the flat \teq{\erg^{-2/3}} form
that signifies curvature or synchrotron emission from
quasi-monoenergetic pairs.  For the Vela pulsar (see Pavlov et al., 2001
for a spectrum), when such a flat slope is extrapolated from the
gamma-ray band, the model flux in the optical band grossly
underpredicts that observed, implying that another component is present
below X-ray wavelengths.  Note that when the surface field becomes
sufficiently high, ground state pair creation suppresses the
synchrotron component and the \teq{\emin} feature disappears.

\begin{figure}[ht] 
\vspace{-90pt} 
\centerline{\hskip 1.3truecm\epsfig{file=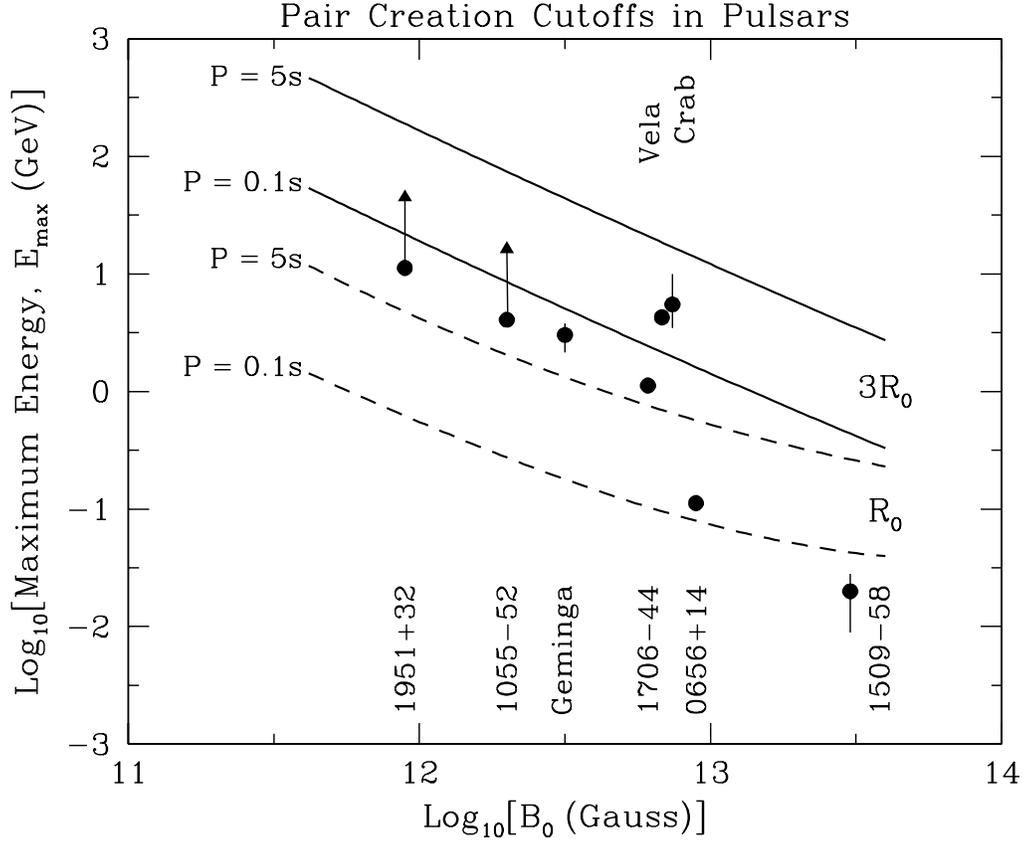, width=0.9\hsize}} 
\vspace{-90pt} 
\caption{
\ Maximum pulsar emission energies (adapted from Baring and Harding, 2000)
imposed by pair creation attenuation at two different altitudes,
\teq{R_0} (dashed curves) and \teq{3R_0} (solid curves), described
empirically via Eq.~(\ref{eq:emax}).  For each altitude, a range
of pulse periods (polar cap sizes) is represented, as indicated.  These
energies are determined by the comprehensive photon
propagation/attenuation code described in Baring and Harding (2001),
which includes curved spacetime effects.  Inferred cutoff energies (or
ranges) for 8 gamma-ray pulsars of different \teq{B_0} are indicated,
from which a trend of declining altitude of emission with increasing
\teq{B_0} is suggested.  Photon splitting will reduce the cutoff
energies below those depicted for \teq{B_0} exceeding around
\teq{0.7B_{\rm cr}}.
} 
\label{fig:hecutoff} 
\end{figure}

The second major spectral feature is the maximum energy of emission,
which is controlled by attenuation due to magnetic pair creation
\teq{\gamma\to e^{\pm}} during photon propagation through the pulsar
magnetosphere.  Such attenuation provides a characteristic
super-exponential turnover (e.g. Daugherty and Harding, 1996) that
contrasts that expected in outer gap models (e.g. see Thompson, 2001, for
a comparison).  Pair creation occurs at the threshold
\teq{\erg\sin\theta_{\rm kB} =2} for high fields, i.e.  \teq{B\gtrsim
0.1 B_{\rm cr}}, and above threshold at \teq{\erg\sin\theta_{\rm kB}
\sim 0.2 B_{\rm cr}/B} for lower fields (e.g. see Daugherty and Harding,
1983).  Here, \teq{\theta_{\rm kB}} is the angle of photon propagation
relative to {\bf B}, and hereafter photon energies \teq{\erg} are
expressed in units of \teq{m_ec^2}.  Hence, the mean free path for
photon attenuation in {\it curved} fields is \teq{\lambda_{\rm pp} \sim
\rho_c/\erg\, \max \{ 2, \; 0.2/B \} }, i.e.  when
\teq{\erg\sin\theta_{\rm kB}} crosses above threshold during
propagation.  The radius of field curvature is \teq{\rho_c =
[Prc/2\pi]^{1/2}} for a pulsar period \teq{P}.  The approximate
dependence of pair creation cutoff energies \teq{\emax} on \teq{B_0},
\teq{R_0} and pulsar period \teq{P} (in seconds) can be summarized in
the relation (Harding, 2001; Baring, 2001a)
\begin{equation} 
   \emax \approx 0.4 \sqrt{P} \, \biggl( \dover{r}{R_0} \biggr)^{1/2} \; 
   \max \Biggl\{ 1,\; \dover{0.1\, B_{\rm cr}}{B_0}\,  
   \biggl( \dover{r}{R_0} \biggr)^3 \Biggr\}\; \hbox{GeV} \;\; . 
 \label{eq:emax} 
\end{equation} 
Accurate numerical determinations derived from the codes developed in
HBG97 and Baring and Harding (2001), are plotted in
Figure~\ref{fig:hecutoff}; these include the effects of general
relativity on spacetime curvature, field enhancement and photon energy
in non-rotating systems.  At fields \teq{B_0\gtrsim 0.7B_{\rm cr}}
photon splitting acts to further reduce \teq{\emax}, as discussed in
Baring and Harding (2001); the operation of splitting proved necessary
to account for the turnover inferred from Comptel data and EGRET upper
limits to PSR 1509-58 (HBG97).  For magnetars, pulsars with fields
above \teq{4\times 10^{13}}Gauss, photon splitting and pair creation
should prohibit any emission above \teq{\sim 100}MeV, though prominent
signals below 100 MeV are possible (Baring, 2001b) due to
the efficiency of resonant Compton scattering.

There is clearly a strong anti-correlation between the maximum energy
and the surface magnetic field, which seems to be augmented by an
apparent decline of emission altitude with \teq{B_0}.  Such a trend is
a distinctive characteristic that can be probed by GLAST; there appears
to be no prediction of such a trend in outer gap models.  The maximum
energy is generally in the 1--10 GeV band for normal young pulsars, but
can be much lower (e.g. HBG97; BH01) for highly magnetized ones, and
also much higher for millisecond pulsars so that sub TeV-band (i.e.
\teq{\sim 50}--100 GeV) signals are possible (Bulik, et al.,
2000) for polar cap models via synchrotron/curvature cascades if the
field is low enough.  It also should be remarked that the cutoff energy
depends on pulse phase, with slightly greater values achieved between
the pulse peaks in the case of Vela modelling (Daugherty and Harding,
1996); such a property matches the EGRET observations (Kanbach et al.,
1994).  Furthermore, there is an asymmetry in \teq{\emax} values
between the peaks due to geometrical effects that is discussed in the
next Section below.  In passing it is noted that it is a generic
property of polar cap models that small rotator obliquities (\teq{<45}
degrees between rotation and dipole axes) are usually needed to
simultaneously produce the observed \teq{\emax} values and double
peaked pulse profiles of phase separation appropriate for gamma-ray
pulsars (Harding and Daugherty, 1998).

The gamma-ray spectral index \teq{\alpha} extending below \teq{\emax}
depends on the details of the cascading, and its behavior can be
broadly summarized as follows.  The primary photon index from cooling
curvature radiation is \teq{5/3}, and should be realized in millisecond
pulsars (Bulik, et al., 2000), where the fields are sufficiently
low that synchrotron components contribute only below 1 MeV.  For
Vela-like fields, successive generations of cascading sequentially
steepen the index (Wei, et al., 1997; Harding and Daugherty, 1998),
saturating at \teq{\alpha =2}.  While most gamma-ray pulsars have
\teq{\alpha < 2} and can be simply described via such cascading, the
Crab possesses a spectrum steeper than \teq{\alpha =2} that must depend
on a spatial convolution in some subtle and obscure manner.  When the
local field in the emission region exceeds around \teq{6\times
10^{12}}Gauss, such as should be the case for PSR 1509-58, pair
creation occurs only in the ground state and quenches cascading,
yielding just a bare, flat curvature spectrum (\teq{\alpha =5/3}).  The
contribution from resonant Compton scattering can also be significant
in high field pulsars, depending on the proximity of the acceleration
region to the surface and the surface X-ray temperature; its extremely
flat spectrum is discussed below in the context of magnetars.

The presence of strong fields virtually guarantees a strong
polarization signal in polar cap models, and when these couple with
spectral structure and temporal information, particularly powerful
observational diagnostics are possible.  This may be fruitful at the
lower end of the cascade continuum in Vela-like objects, but it is an
especially valuable tool for highly-magnetized pulsars since the
attenuation cutoffs fall in the Comptel band, and should exhibit strong
and distinctive polarization signatures.  Hard gamma-ray experiments
like GLAST are generally not afforded the opportunity to act as
polarimeters, being limited by multiple scattering in trackers above
300 MeV.  Medium energy gamma-ray experiments, on the other hand, are
ideally suited to polarization studies (via their sampling of Compton
scattering kinematics).  Gamma-ray polarimetry is no longer a distant
dream, given the prospects (Lei, et al., 1997) that the
recently-launched INTEGRAL mission will detect polarization at the 10\%
level from the Crab pulsar (at 200--600 keV), and also in a handful of
other sources.  Polarimetric capability in the hard X-ray and soft
gamma-ray bands is a high priority for next-generation advanced Compton
telescopes (e.g. see Kanbach et al., 2000, and the Web pages for the
MEGA [{\tt http://www.gamma.mpe-garching.mpg.de/MEGA/mega.html}] and
ACT [{\tt http://gamma.nrl.navy.mil/ngram/}] consortia).

\subsection*{Observabilities}

Spectral signatures provide the second layer of detail in observational
diagnostics of gamma-ray pulsars; the primary layer is clearly the flux
of a source at earth, i.e. its observability.  This essentially
represents the normalization of the model spectrum convolved with the
pulsar's distance.  In recent years there has been an evolution in the
understanding of what controls the detectability, making way for a
further dimension of discrimination between polar cap and outer gap
models in the GLAST era, when luminosity trends will not be crimped by
small population statistics.  The traditional approach of the EGRET
community was to use the dipole spin-down power \teq{{\dot E} \propto
B_0^2/P^4} as an indicator of a pulsar's observability, which is in
fact mirrored by the population of field X-ray pulsars (Becker and 
Tr\"umper, 1997); recent Chandra detections of pulsars in globular
clusters indicate (Grindlay et al., 2002) a weaker luminosity dependence
on \teq{B_0/P^2}.  While theoretically motivated (see just below), the
spin-down luminosity choice did not match the trend subsequently
established by EGRET: that there is clearly an almost linear
correlation  between the inferred luminosity \teq{L_{\gamma}} of
gamma-ray pulsars and \teq{B_0/P^2}, i.e. the voltage drop across the
open field lines for standard polar caps.  There is a very modest
scatter in this correlation, largely due to the uncertainty in
determining source distance by folding radio dispersion measures into
the Taylor and Cordes (1993) galactic electron model (just recently 
updated in Cordes and Lazio, 2002).

It is interesting to note that such a linear correlation with
\teq{B_0/P^2} of the inferred luminosity of gamma-ray pulsars was {\it
predicted} (Harding, 1981) in the context of the polar cap model.  At
the time, only 2 gamma-ray pulsars were known, strengthening the impact
of the polar cap model on the understanding of magnetospheric emission
in such pulsars.  The contention of a \teq{L\propto B_0/P^2} dependence
is based on the premises that the radiative luminosity is proportional
to the Goldreich-Julian current, and that the cascade emission is
initiated by pairs of Lorentz factor that is almost independent of
pulsar \teq{B_0} and period.  Subsequent predictions by competing
analyses/models (e.g. Sturner and Dermer, 1994; Romani and Yadigaroglu,
1995; Cheng et al., 1998; Rudak and Dyks, 1999) and revisions (Zhang and 
Harding, 2000) all post-dated the EGRET database.  The current status
is that the polar cap expectations (Sturner and Dermer, 1994; Zhang and 
Harding, 2000) match the data slightly more accurately than their outer
gap counterparts (Romani and Yadigaroglu, 1995; Cheng et al., 1998),
with each group of researchers offering different \teq{B_0} and \teq{P}
dependences for the luminosity (see Harding, 2001 for a review).  This
situation is presently limited by small number statistics, however in
the GLAST era such correlations will be established on a firm basis.
 
Motivations for considering their luminosity dependence are not
confined to model discrimination and refinement; assumed luminosity
``laws'' can dictate period selection in pulsation searches.  This is a
salient issue for GLAST, since it will be capable of blind period
searches on gamma-ray sources with no radio counterparts.  The period
dependence is the most critical element to the \teq{L_{\gamma}(P,\,
{\dot P})} relationship.  While EGRET observed most pulsars high up on
a \teq{B_0^2/P^4/\dpsrsq} rank-ordered list (where \teq{\dpsr} is the
pulsar distance), certain gamma-ray pulsars (notably the longer period
pulsars PSR 0656+14 and PSR 1055-52) are surprisingly low in spin-down
luminosity, and millisecond pulsars proved extraordinarily difficult to
detect, till the detection of PSR 0218+4232 (see Kuiper et al., 2000).
Clearly, a gamma-ray luminosity dependence \teq{L_{\gamma}(P,\, {\dot
P})} that differs from the spin-down one will dramatically modify the
observability criterion, particularly if the period dependence is
substantially different.  Accurate determination of
\teq{L_{\gamma}(P,\, {\dot P})} will influence the weight on different
period ranges that will be applied to GLAST source data in pulsation
searches.  This will be particularly germane to cases where the source
is near the galactic plane and pulsations are not evident at the
frequencies of known radio pulsars within the GLAST source
localization.

The  spectral shape also affects the observability (Baring and Harding,
2000), a more subtle influence.  This is a consequence of how the
luminosity is distributed in the gamma-ray band, specifically that
portion that emerges above the threshold sensitivity for a specific
gamma-ray detector.  The driving parameter for such an apportionment is
the maximum emergent energy \teq{\emax}, and to a lesser extent
\teq{\emin} and the gamma-ray spectral index \teq{\alpha}, since the
spectra are generally flat enough for the bulk of the luminosity to
emerge at the highest energies.  These parameters control the
normalization of the pulsar gamma-ray power-law for a given
luminosity.  An appropriate definition of a detector's observability
\teq{{\cal O}(\ethresh )} is the {\it integral flux} above an effective
instrumental energy threshold \teq{\ethresh}.  For pulsars with
\teq{\emin\ll\ethresh}, the usual case for GLAST considerations, this
scales as the luminosity divided by the spectral normalization,
yielding \teq{{\cal O}(\ethresh )\propto L_{\gamma}(P,\, {\dot P})\,
\emax^{(\alpha -2)}/\dpsrsq} (Baring and Harding, 2000; Baring, 2001a).
Modest modifications to this dependence are possible, in particular if
\teq{\emin\gtrsim \ethresh} (see Baring 2001a for a discussion).  The
essential feature is that the observability is substantially different
from that inferred from the spin-down formula, assuming that
\teq{\emax} scales with \teq{B_0} and \teq{P} approximately as the low
field alternative offered in Eq.~(\ref{eq:emax}).  Accordingly,
observabilities predicted for GLAST pulsation searches should follow a
dependence (Baring, 2001a) somewhere in between \teq{B_0/P^2} and
\teq{B_0^2/P^{5/2}}.  Using the latter possibility, Baring and Harding
(2000) generated a revised rank-ordered listing that indicated a
dramatic rearrangement from the traditional EGRET ordering.  A
depiction of this re-ordering in \teq{{\cal O}(\ethresh )} versus
\teq{{\dot E}/\dpsrsq} space is given in Figure 2, where each axis can
be used to define a rank ordering.

The limited scatter around a linear dependence in the figure indicates
that there are generally only modest changes to the rankings for most
pulsars when adopting the observability as an updated criterion for
detectability.  The most notable changes (see Baring, 2001a) included the
much higher ranking of the ``outlier'' longer period gamma-ray pulsars
PSR 0656+14 and PSR 1055-52, and the marked lowering of millisecond
pulsars (PSR 1939+2134, PSR 0437-4715, PSR 1744-1134, etc.) in the
ranks, specifically out of the top 40.  Both of these reflect the
weaker dependence of the revised observability on \teq{P}.  These
refinements of rank-orderings mute questions of why PSR 1055-52 was
seen by EGRET.  Further revisions are in progress, and largely focus on
the details of spectral shape.  Only modest influences are expected
from new pulsar distance determinations resulting from the updated
model for the galactic electron distribution, NE2001 (Cordes and Lazio,
2002).

\begin{figure}[ht] 
\vspace{-10pt} 
\centerline{\hskip 5.6truecm\epsfig{file=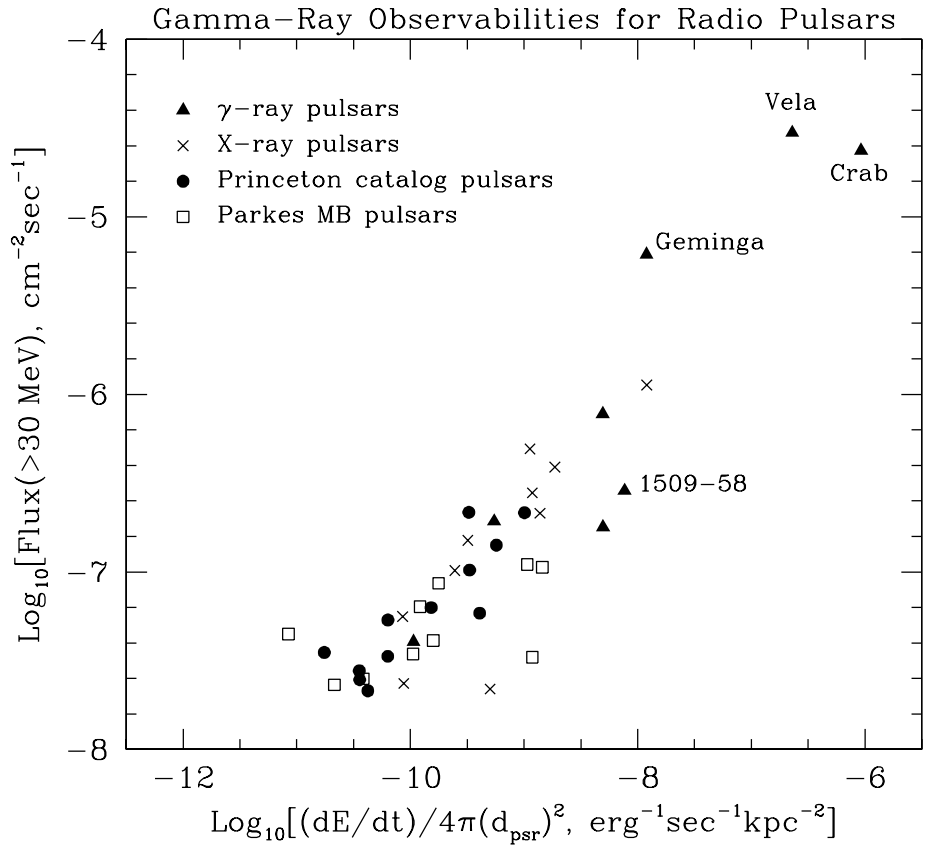, width=0.7\hsize}} 
\vspace{-20pt} 
\caption{
\ The GLAST observability phase space for 40 known radio pulsars that 
are grouped by symbol, as indicated, according to whether they are
gamma-ray pulsars, X-ray emitters, ordinary Princeton catalog
sources, or new pulsars from the Parkes MultiBeam (MB) survey 
[Manchester et al., 2001; {\tt http://www.atnf.csiro.au/research/pulsar/pmsurv/}].
The x-axis represents the canonical dipole spin-down power \teq{{\dot
E}} divided by \teq{\dpsrsq}, and the y-axis indicates the fiducial
GLAST observability \teq{{\cal O}(30 MeV)}, i.e. integral flux above
\teq{30} MeV, as obtained by Baring and Harding (2000).  The GLAST
integral flux threshold for pulsating sources is around \teq{2\times
10^{-9}} cm$^{-2}$sec$^{-1}$, so that it should comfortably detect all
depicted pulsars.  The corresponding EGRET value is \teq{3\times
10^{-7}} cm$^{-2}$sec$^{-1}$, above around 100 MeV.  Observe that the
Parkes MB pulsars do not rank in the top ten for flux due to their
typically large distances.  Also, the flux specification for the high
field pulsar PSR 1509-58 is sensitive to fine-tuning in \teq{\emax},
since this parameter is proximate to the 30 MeV threshold for the GLAST
Large Area Telescope.
} 
\label{fig:glastrank} 
\end{figure}

\section*{RECENT DEVELOPMENTS}  

\section*{Slot Gaps}
 \label{sec:slotgaps}

During the Compton Gamma-Ray Observatory era, polar cap modelling
focussed mostly on magnetospheric emission processes in order to take
advantage of the rapid increase in gamma-ray pulsar data afforded by
EGRET.  In the last few years, considerable effort has been invested by
Harding and Muslimov in the exploration of the acceleration region in
order to determine the altitudes and particle energies associated with
acceleration in the polar cap potentials (e.g. see Harding and Muslimov
1998, Muslimov and Harding 2003).  Such details are essential to forming 
a self-consistent
polar cap model for outward cascade emission as well as for polar cap
reheating and subsequent X-ray reprocessing of the cascade power.  In
particular, a major question was raised by the modeling of Vela data:
Daugherty and Harding (1996) ascertained that high altitudes (\teq{\sim
2-3} stellar radii) for the emission region were required to
simultaneously explain the pulse phase separation and the \teq{\sim
5}GeV maximum energy of emission.  The altitude of the pair formation
front (PFF) was a free parameter in their model, and a concerted effort
in studying pulsar electrodynamics was needed to ascertain whether high
altitudes could naturally be expected.

The recent work of Muslimov and Harding (2003) has affirmed the altitude
choice of Daugherty and Harding (1996).  In an extension of their
sequence of papers on PFF development and location under the influence
of general relativistic frame-dragging effects, they have explored
colatitudinal variations of the PFF height.  Their principal result is
that there is a large variation in height with colatitude \teq{\theta}
along the surface of the polar cap.  This arises because the
accelerating potential drops smoothly to zero at the rim of the polar
cap, thereby extending the acceleration region to high altitudes in the
vicinity of the rim.  Pair quenching of the smaller electric field is
precipitated more gradually due to the prolongation of acceleration in
this confined region, referred to as the {\it slot gap} (after Arons,
1983).  The result is that the dominant emission region spans a range
of altitudes from relatively near the surface at the magnetic axis up
to several stellar radii above the surface in the vicinity of the rim.
Accordingly, hard X-ray and gamma-ray pulsar emission contains both
core and conal components.  In addition, the broad distribution of
emission altitudes could aid smoothing of the super-exponential
turnovers imposed by pair creation attenuation near \teq{\emax}.  The
angular width of the slot gap scales approximately as \teq{(P/B)^{1/2}}
for \teq{B_0\lesssim 4\times 10^{12}}Gauss.  When this is incorporated
in solid angle modifications to the emergent gamma-ray luminosity, the
slot gap model of Muslimov and Harding (2003) impressively accounts for
the inferred luminosities in most cases; Geminga and the millisecond
pulsar PSR 0218+4232 provide exceptions that require refinements to the
model.

\section*{Pulse Asymmetry}

The GLAST experiment will afford phase-resolved spectroscopy at
unprecedented statistical significance.  It will provide well-defined
pulse profiles in much smaller energy bins than were possible with
EGRET or Comptel.  Such developments enable new diagnostic
capabilities.  One effect that can be probed by this advance is that of
pulse asymmetry, which has been studied in detail by Dyks and Rudak
(2002).  As the magnetosphere rotates, the photon trajectories slip
across the curved field lines slightly.  This relative motion differs
between the leading and trailing rims of the polar cap, translating to
different shapes for the two peaks in the pulse profile.  Since pair
creation rates are critically sensitive to the angle \teq{\theta_{\rm
kB}} between the photon path and the local field line, one expects
significant correlation of pulse asymmetries with photon energy.  Dyks
and Rudak find that near the pair creation turnover at \teq{\emax}, the
trailing peak dominates since \teq{\theta_{\rm kB}} is larger on
average for the leading rim.  Equivalently, \teq{\emax} is lower for
the leading pulse.  Concomitantly, the pair reprocessing is enhanced
for the leading peak so that it becomes the more prominent of the two
peaks at sub-GeV energies.   While \teq{\emax} is an increasing
function of \teq{P} for the leading rim, non-monotonic dependence of
\teq{\emax} on period is exhibited by the trailing rim, providing
constraints that will aid isolation of the rotator obliquity for an
assumed polar cap size.  Note that although Dyks and Rudak explored
these asymmetry effects in flat spacetime, they are also present in the
curved spacetime magnetospheric models of Muslimov and Harding (2003).
Observe also that this rotational effect is a strong function of pulse
period, being much more pronounced for fast rotators, as expected.
Fortunately, these are the brighter portion of pulsars so such
distinctive properties will provide powerful probes of polar cap models
in the GLAST era.

\section*{POWERFUL OBSERVATIONAL DIAGNOSTICS}

In conclusion, this paper has identified key properties that the polar
cap (PC) model has that are dependent on high \teq{B} physics, and that
are palpably distinguishable from outer gap model characteristics;
these can be summarized as follows.  In the PC model there is no pulsed
TeV emission in garden-variety pulsars, though millisecond pulsars
should exhibit sub-TeV emission.  The attenuation of gamma-rays due to
one photon pair creation generates super-exponential cutoffs in the 10
MeV -- 10 GeV band when the surface polar field is in the range
\teq{10^{11}}Gauss \teq{\lesssim B_0\lesssim 4\times 10^{13}}Gauss;
these have been posited as a key pulsar diagnostic for GLAST.
Moreover, \teq{\emax} should decline with increasing \teq{B_0} if
\teq{B_0\lesssim 10^{14}}Gauss.  At the same time, an increase
of \teq{\emax} with emission altitude implies a correlation between
\teq{\emax} and pulse separation.  Recent work has also suggested that
there should exist a significant pulse asymmetry near \teq{\emax} and
for lower cascade photon energies.  In the special case of magnetars,
pulsars with fields above \teq{4\times 10^{13}}Gauss, attenuation dues
to photon splitting and pair creation should prohibit any emission
above \teq{\sim 100}MeV, though prominent sub-100 MeV signals
are possible due to the efficiency of resonant Compton scattering, and
can be probed by GLAST.  The hard X-ray/soft $\gamma$-ray spectral
slope is expected to steepen slightly as \teq{B_0} rises when
\teq{3\times 10^{11}}Gauss \teq{\lesssim B_0\lesssim 10^{13}}Gauss,
though flat curvature radiation spectra with \teq{\alpha =5/3} are
anticipated for both highly-magnetized and millisecond pulsars, i.e.
at both ends of the pulsar field range.  Most of these features will be
probed by new gamma-ray missions INTEGRAL, AGILE, GLAST and
ground-based air \v{C}erenkov telescopes (HESS, Veritas, MAGIC,
CANGAROO-III), as well as current X-ray missions CHANDRA, XMM, and
RXTE.  In addition, strong and distinctive polarization signatures are
expected in pulsar spectra, which may be explored for the first time by
INTEGRAL; positive detections anticipated for the Crab pulsar would
break new ground in for the astrophysics of gamma-ray pulsars.


\vskip 10pt
\noindent
E-mail address of Matthew G. Baring: baring@rice.edu

\end{document}